\UseRawInputEncoding
\pdfoutput=1
\iftrue
\documentclass[aps,prl,twocolumn, 
			   groupedaddress,superscriptaddress,
			   amsfonts,amssymb,amsmath,
			   citeautoscript,
			   a4paper]{revtex4-1}
\else
\documentclass[aps,pra,preprint,groupedaddress,superscriptaddress,notitlepage,
			   amsfonts,amssymb,amsmath,
			   citeautoscript,
			   a4paper]{revtex4-1}
\fi

\usepackage[pdftex]{hyperref}
\hypersetup{colorlinks,
			linkcolor={blue!75!black!80!yellow},
			citecolor={blue!75!black!80!yellow},
			urlcolor={blue!75!black!80!yellow},
			pdfstartview=FitH}
\usepackage{graphicx}
\usepackage{mathrsfs}
\usepackage{amsthm}
\usepackage{xspace}
\usepackage{braket}
\usepackage{xr}
\usepackage{gensymb} 
\usepackage{xcolor,soul}
\usepackage{stmaryrd} 
\usepackage[UKenglish]{babel}
\usepackage{placeins} 
\usepackage{siunitx}
\DeclareSIUnit[number-unit-product=]\percent{\char`\%} 
\usepackage{makecell}
\usepackage{diagbox}
\usepackage{float}
\usepackage{amsmath} 

\usepackage{txfonts}  
\usepackage{txfontsb} 
\usepackage{extarrows}

\newcommand{\appropto}{\mathrel{\vcenter{
			\offinterlineskip\halign{\hfil$##$\cr
				\propto\cr\noalign{\kern2pt}\sim\cr\noalign{\kern-2pt}}}}}

\newcommand{\eg}{e.g.\@\xspace}

\newcommand{\kns}{\mathbf{k}\!-\!{\mathrm{NS}}}
\newcommand{\gx}{\gamma_x}
\newcommand{\gy}{\gamma_y}
\newcommand{\lx}{\lambda_x}
\newcommand{\ly}{\lambda_y}

\makeatletter
\newcommand*{\addFileDependency}[1]{
  \typeout{(#1)}
  \@addtofilelist{#1}
  \IfFileExists{#1}{}{\typeout{No file #1.}}
}
\makeatother

\usepackage{scalerel}

\usepackage{textcomp} 
\usepackage{xifthen}
\usepackage{etoolbox}
\newboolean{togglecomments}
\newboolean{togglechanges} 

\setboolean{togglecomments}{true}
\setboolean{togglechanges}{false}

\newcommand{\comment}[2]{%
    \ifbool{togglecomments}%
    {\textcolor{blue!70!black}{\small\textsf{%
    \textsuperscript{\textsc{\textsf{\MakeLowercase{#1}}}}%
    [#2]}}} 
    {}}     
\newcommand{\swap}[2]{\ifbool{togglechanges}
    {#2}  
    {\textcolor{red!70!black}{[#1]}\textrightarrow{}\textcolor{green!50!black}{[#2]}}}
\newcommand{\remove}[1]{\ifbool{togglechanges}
    {}    
    {\textcolor{red!70!black}{#1}}}
\newcommand{\inset}[1]{\ifbool{togglechanges}
    {#1}  
    {\textcolor{green!50!black}{#1}}}
\newcommand{\optional}[1]{\ifbool{togglechanges}
    {#1}  
    {\textcolor{yellow!50!orange!80!gray}{#1}}}
\newcommand{\citeremind}[1]{%
    [\textcolor{blue!75!black!80!yellow}{
        $\blacksquare$%
           \ifthenelse{\isempty{#1}}
               {}
               {\textsuperscript{\textsf{#1}}}%
        }]\xspace}
\newcommand{\todo}[1]{
    \textcolor{orange!80!yellow!95!black}{\textbf{[}%
        \ifthenelse{\isempty{#1}}%
        {\text{$\blacksquare$}}%
        {{\small\textsf{#1}}}%
        \textbf{]}}}
        
\newcommand{\hkuaffil}{\footnotesize Department of Physics, The University of Hong Kong, Pokfulam, Hong Kong, China}
\newcommand{\usstaffil}{\footnotesize College of Optical-Electrical Information and Computer Engineering,
University of Shanghai for Science and Technology, Shanghai 200093, China}

\begin{document}

\title{Synthetic gauge fields enable high-order topology on Brillouin real projective plane}

\author{Jinbing Hu}
\affiliation{\hkuaffil}
\affiliation{\usstaffil}
\author{Songlin Zhuang}
\affiliation{\usstaffil}
\author{Yi~Yang}
\email{yiyg@hku.hk}
\affiliation{\hkuaffil}

\begin{abstract}
The topology of the Brillouin zone, foundational in topological physics, is always assumed to be a torus. We theoretically report the construction of Brillouin real projective plane ($\mathrm{RP}^2$) and the appearance of quadrupole insulating phase, which are enabled by momentum-space nonsymmorphic symmetries stemming from $\mathbb{Z}_2$ synthetic gauge fields.
We show that the momentum-space nonsymmorphic symmetries quantize bulk polarization and Wannier-sector polarization nonlocally across different momenta, resulting in quantized corner charges and an isotropic binary bulk quadrupole phase diagram, where the phase transition is triggered by a bulk energy gap closing.
Under open boundary conditions, the nontrivial bulk quadrupole phase manifests either trivial or nontrivial edge polarization, resulting from the violation of momentum-space nonsymmorphic symmetries under lattice termination.
We present a concrete design for the $\mathrm{RP}^2$ quadrupole insulator based on acoustic resonator arrays and discuss its feasibility in optics, mechanics, and electrical circuits. 
Our results show that deforming the Brillouin manifold creates opportunities for realizing high-order band topology. 
\end{abstract}

\maketitle

The topology of Brillouin zone (BZ) is widely known as a torus, around which the state evolution encodes the topology of solids and artificial materials.
However, broadly speaking, a torus is only one closed compact manifold and can thus be transformed into other manifolds; this transition should lead to rich topological consequences.
To this end, the Brillouin torus was recently transformed into a Klein bottle under a momentum-space non-symmorphic ($\kns$) symmetry that performs glide reflection in momentum space~\cite{chen2022brillouin}.
In contrast to the conventional nonsymmorphic symmetries that act on real space and feature symmetry-invariant momenta (\eg Refs.~\cite{wieder2018wallpaper,lu2016symmetry,armitage2018weyl,zhao2016nonsymmorphic,wieder2016double,liu2014topological,fang2015new}), $\kns$ symmetries act on momentum space and symmetry invariant momenta may not be identified~\cite{chen2022brillouin}.
The key to creating such $\kns$ symmetries is synthetic gauge fields~\cite{aidelsburger2018artificial}, which can projectively modify point-group symmetries into space-group ones~\cite{zhao2020z,chen2022brillouin,chen2023classification}.
Beyond the torus, only the Brillouin Klein bottle was synthesized under a single $\kns$ symmetry so far~\cite{chen2022brillouin}, leading to interesting nonlocal first-order topology.

The electric polarization of crystals \cite{resta1992theory,king1993theory,vanderbilt2018berry} was recently extended to higher electric multipole moments, giving rise to topological multipole insulators characterized by high-order band topology~\cite{benalcazar2017quantized,benalcazar2017electric,schindler2018higher}. 
Although first proposed in tight-binding models, topological quadrupole insulators have been observed in various physical platforms (\eg Refs.~\cite{peterson2018quantized,mittal2019photonic,qi2020acoustic,ni2020demonstration,xue2020observation,serra2018observation,imhof2018topolectrical}). 
In addition, the topological quadrupole phase was also observed in nonsymmorphic systems, where quantized quadrupole moment is protected by real-space nonsymmorphic reflections~\cite{lin2020anomalous,zhang2020symmetry,liu2023second,mao2022third,liu2022geometric}.
However, high-order band topology on manifolds beyond the Brillouin torus has not been investigated yet.

In this work, we theoretically introduce a quadrupole topological insulator on a two-dimensional Brillouin real projective plane ($\mathrm{RP}^2$) via synthetic gauge fields.
Two $\kns$ symmetries, emerged from $Z_2$ gauge fields and the associated checkerboard flux pattern, partition the original BZ torus into several non-unique $\mathrm{RP}^2$ manifolds, and any of them can be equivalently used for topology diagnosis.
The symmetries of $\kns$ lead to the quantization of bulk and Wannier-sector polarization nonlocally across different momenta. An isotropic binary bulk phase diagram featured by bulk energy gap closing is obtained, in contrast to the quadrupole insulator associated with Wannier band closing~\cite{benalcazar2017quantized,peterson2018quantized,mittal2019photonic,qi2020acoustic,serra2018observation,imhof2018topolectrical}.
For a nontrivial bulk quadrupole protected by $\kns$ symmetries, the edge polarizations can still be trivial or non-trivial since lattice termination unavoidably violates the $\kns$ symmetry, indicating a breakdown of the bulk-boundary correspondence.
Finally, we present a concrete design of the $\mathrm{RP}^2$ quadrupole insulator using acoustics and discuss other possible realizations in optics, mechanics, and circuits.

\begin{figure}[htbp]
    \centering
    \includegraphics[width=\linewidth]{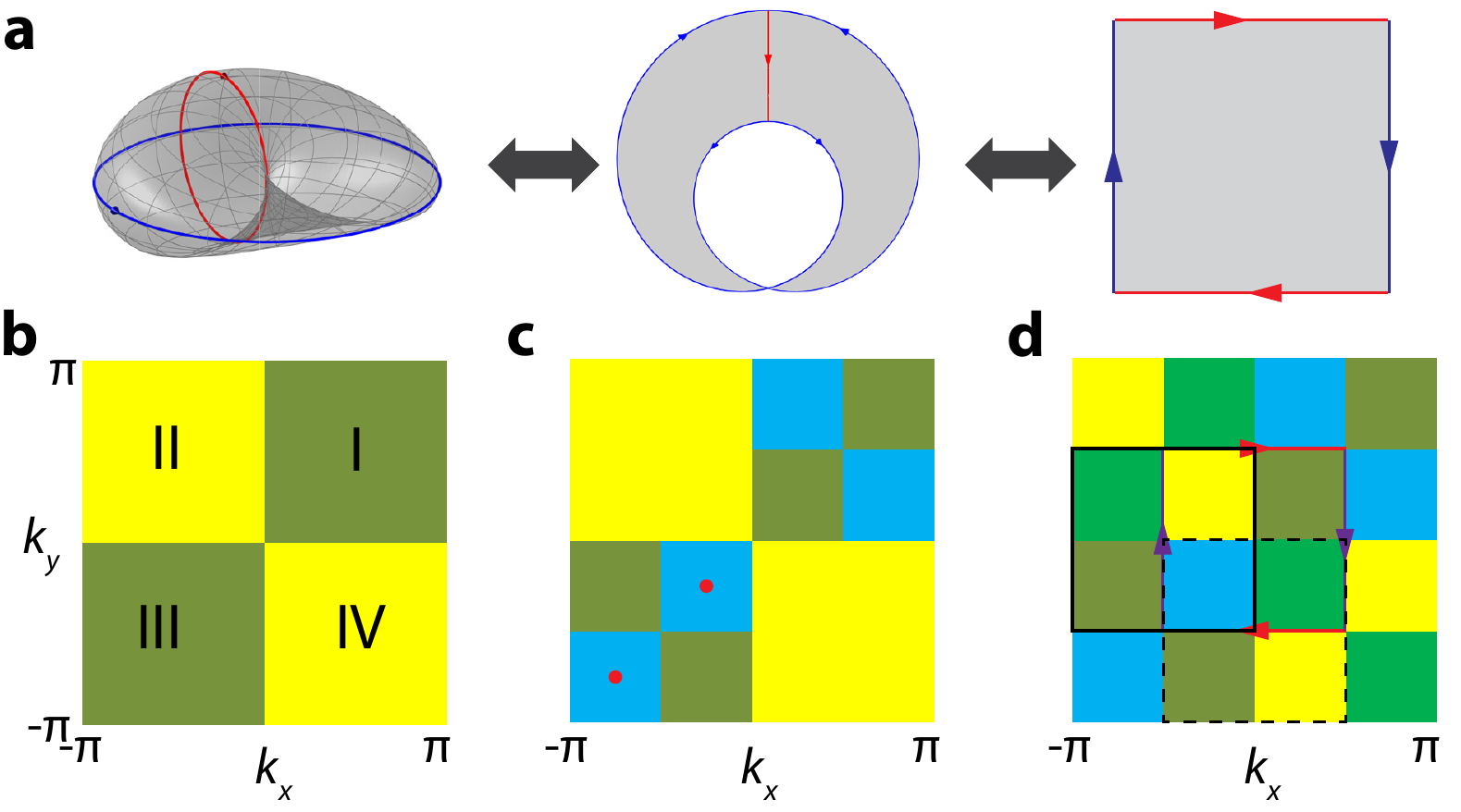}
    \caption{\textbf{Real projective plane in momentum space}. 
                \textbf{a.} Folding a real projective plane (left) from a rectangle (right), whose opposite sides are glued in an anti-parallel manner (denoted by color arrows). 
                \textbf{b.} The $\kns$ mirror symmetries $M_x$ and $M_y$ divide the first Brillouin zone (BZ) $[-\pi,\pi)\times[-\pi,\pi)$ into four quadrants I-IV.
                \textbf{c.} The $\kns$ inversion $P_{\pi}$ further divides each quadrant into diagonal and anti-diagonal partners (\eg green and blue colors in the I and III quadrants. The red dots indicate two momenta linked by $P_\pi$.
                \textbf{d.} $M_x$, $M_y$ and $P_{\pi}$ jointly divide the first BZ into sixteen plaquettes. A reduced BZ can be defined using any four plaquettes of distinct colors, \eg the square at the zone center surrounded by red and violet arrows. Other examples include the squares in black solid and dashed lines.
                }                
    \label{fig1}
\end{figure}

\textit{Brillouin real projective plane.}
$\mathrm{RP}^2$ is a two-dimensional surface generalizing the M\"{o}bius strip. It is constructed from a square by identifying two pairs of opposite edges with a half twist (Fig.~\ref{fig1}a).
Mathematically, the construction of $\mathrm{RP}^2$ can be represented as a unit square ($[0,1]\times[0,1]$) with opposite edges following the equivalence relations (see Sec. I of \cite{SM_note})
\begin{subequations}
    \label{Eq1}
    \begin{align}
        (x,0) &\sim (1-x,1)~~\mathrm{for}~~0 \leq x \leq 1,\\
        (0,y) &\sim (1,1-y)~~\mathrm{for}~~0 \leq y \leq 1.
    \end{align}
\end{subequations}
We aim to realize these conditions to realize $\mathrm{RP}^2$ in momentum space.
The $\kns$ reflections $M_x$ and $M_y$ meet the need, as they constrain the wave vector $\textbf{k}=(k_x,k_y)$ via
\begin{subequations}
        \begin{align}
            M_x: (k_x,k_y)&\rightarrow (-k_x,k_y+\pi) , \label{eq:Mx}\\
            M_y: (k_x,k_y)&\rightarrow (k_x+\pi,-k_y), \label{eq:My}  
        \end{align}
    \label{Eq2}
\end{subequations}
which realizes Eq.~\ref{Eq1} for $(k_x,k_y)\in[-\pi/2,\pi/2]\times[-\pi/2,\pi/2]$ (see Sec. II of \cite{SM_note}). In addition, the simultaneous presence of $M_x$ and $M_y$ naturally defines a $\kns$ inversion $P_{\pi}\equiv M_xM_y$, which confines the wavevectors as
\begin{align}\label{Eq3}
    P_{\pi}: (k_x,k_y) \rightarrow (\pi-k_x,\pi-k_y).
\end{align}
$M_x$, $M_y$, and $P_{\pi}$ further partition the Brillouin torus. Specifically, $M_x$ and $M_y$ divides the first BZ ($[-\pi,\pi]\times[-\pi,\pi]$) into four quadrants (Fig.~\ref{fig1}b). Further, each quadrant is divided into two pairs by $P_{\pi}$: the diagonal and off-diagonal pairs, and each pair is linked by $P_{\pi}$ (Fig.~\ref{fig1}c). Therefore, $M_x$, $M_y$, and $P_{\pi}$ jointly divide the first BZ is divided into sixteen plaquettes (Fig.~\ref{fig1}d) (see Sec. II of \cite{SM_note}). A reduced BZ can be defined on any four plaquettes of distinct colors, \eg the four plaquettes $(k_x,k_y)\in[-\pi/2,\pi/2]\times[-\pi/2,\pi/2]$ at the zone center in Fig.~\ref{fig1}d. 
Notably, the boundaries of the reduced BZ connect in an anti-parallel manner via $M_x$ (red arrows) and $M_y$ (violet arrows), thereby realizing an $\mathrm{RP}^2$ in momentum space. 
There are three remarkable features of the $\kns$ symmetry group. First, analogous to their real-space counterparts, the $\kns$ symmetries exhibit rich Abelian and non-Abelian algebra depending on the (anti-)commutation between $M_x$ and $M_y$~\cite{wieder2018wallpaper,lu2016symmetry,zhang2020symmetry,wang2016hourglass,yang2022non}. Second, the choice of the reduced $\mathrm{RP}^2$ BZ is non-unique (\eg squares in black solid and dashed lines in Fig.~\ref{fig1}d,  as long as four plaquettes in distinct colors are enclosed). Third, the reduced $\mathrm{RP}^2$ BZ remains a closed manifold, enabling adiabatic calculation of topological invariants. 

\begin{figure}
    \centering
    \includegraphics[width=\linewidth]{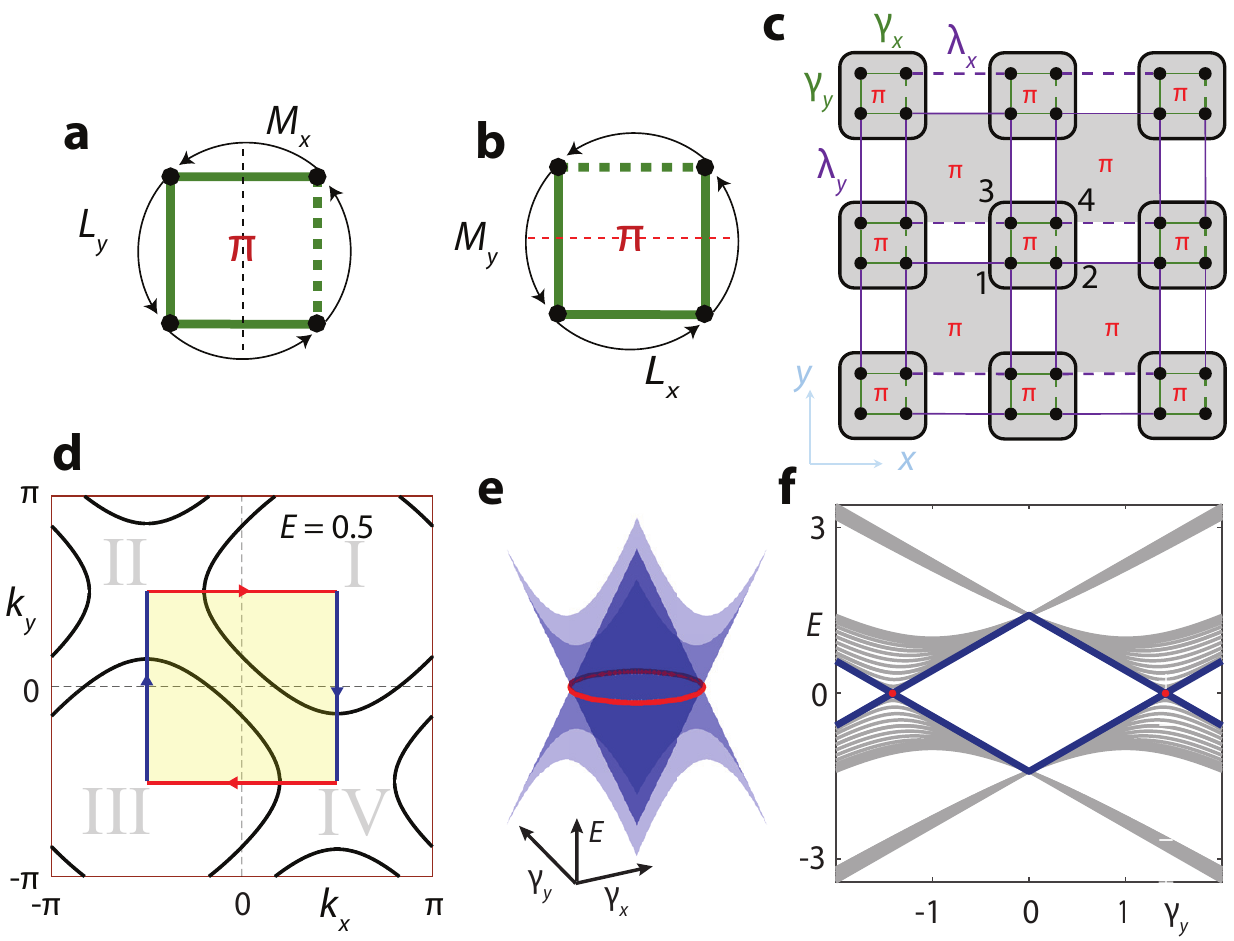}
    \caption{\textbf{Lattice model featuring $\kns$ reflections $M_x$ and $M_y$}.
                \textbf{a-b.} Threading $\pi$ flux projectively modify the conventional reflections $m_i~(i=x,y)$ into $\kns$ ones $M_i~(i=x,y)$, which anticommute with the translational symmetries $L_j~(j=x,y)$ along the other direction.                 
                \textbf{c.} Model with checkerboard $\pi$ flux enables anti-commutative $M_x$ and $M_y$. Solid and dashed lines indicate real positive and negative hoppings.                 
                \textbf{d.} Iso-energy contour plot at $E=0.5$ for $\gx=0.75$ and $\gy=0.68$. 
                \textbf{e.} The middle two bands in $(\gx,\gy)$ space; red circle indicates gap closing. 
                \textbf{f.} Bulk spectrum as a function of $\gy$ for $\gx=0$ with red dots showing gap closing at $\gy=\pm\sqrt{2}$.              
                }
    \label{fig2}
\end{figure}

\textit{Tight-binding implementation.}
Consider the square lattice of four sites (Fig.~\ref{fig2}a), the presence of $\pi$ flux enables an anti-commutative relation between reflection $M_x$ and the translation operator $L_y$ along $y$, i.e., $M_xL_yM_x^{-1}L_y^{-1}=-1$. The right-hand negative sign can be reformulated as a half translation along $k_y$, i.e., $-1=e^{iG_ya/2}$  (where $a$ is the lattice constant and $G_y$ the reciprocal lattice vector along $y$) \cite{chen2022brillouin}. Therefore, besides reversing $k_x$, $M_x$ also contains a half translation along $k_y$, fulfilling Eq.~\eqref{eq:Mx}. Similarly, the relation Eq.~\eqref{eq:My} can be implemented by adding the $\pi$ flux via a different gauge connection in Fig.~\ref{fig2}b. 
Jointly, we arrange the negative hopping along both $x$ and $y$ directions (Fig.~\ref{fig2}a-b) to realize a checkerboard flux pattern (Fig.~\ref{fig2}c) featuring $\kns$ $M_x$ and $M_y$.
The associated spinless Bloch Hamiltonian is
\begin{align}\label{Eq4}
    H(k_x,k_y)=\left[ \begin{array}{cccc}    
                     0 & a_+ & b_+ & 0\\
                     a_+^* & 0 & 0 & b_-\\
                     b_+^* & 0 & 0 & a_-\\
                     0 & b_-^* & a_-^* & 0\\
                     \end{array}
              \right],
\end{align}
with $a_{\pm}=\gx\pm \lx e^{-ik_x}$, $b_{\pm}=\pm \gy+\ly e^{-ik_y}$, and $*$ denoting complex conjugate. Here, $\gx$ and $\gy$ ($\lx$ and $\ly$) represent the intra-cell (inter-cell) hopping amplitudes along $x$ and $y$, respectively. Without loss of generality, we set unity lattice constants and $\lx=\ly=1$.

Under the link arrangements in Fig.~\ref{fig2}c, the $\kns$ symmetries satisfying Eqs.~\eqref{Eq2} and \eqref{Eq3} have the form: $M_x=\sigma_3\otimes\tau_1$, $M_y=\sigma_1\otimes\tau_0$, and $P_{\pi}=i\sigma_2\otimes\tau_1$, together generating a non-Abelian group $D_8$. Here, $\sigma$'s and $\tau$'s are Pauli matrices acting on sites along $y$ and $x$, respectively. 
An example band structure validates the BZ partition in Fig.~\ref{fig1}d, where the band in the reduced $\mathrm{RP}^2$ BZ can be extended to generate that of the full BZ (see Sec.~III of \cite{SM_note}). Such shrinking of the BZ results from the gauge transformation induced by the negative coupling, which doubles the lattice constant along $x$ and $y$ (see Sec. IV of \cite{SM_note}). 

The model is an insulator at zero energy unless the bulk gap closing condition  
\begin{align}
    \sqrt{\gx^2+\gy^2}=\sqrt{\lx^2+\ly^2}
    \label{eq:gap}
\end{align}
is satisfied, as shown in Fig.~\ref{fig2}e and its surface cut at $\lambda_x=0$ in Fig.~\ref{fig2}f [see Sec. III for a detailed derivation of Eq.~\eqref{eq:gap}]. %
The isotropic feature of Eq.~\eqref{eq:gap} implies that the bulk-energy gap closing condition is non-separable in the intra- and inter-cell hopping, which is helpful for understanding the quadrupole phase below. 

\begin{figure}
    \centering
    \includegraphics[width=\linewidth]{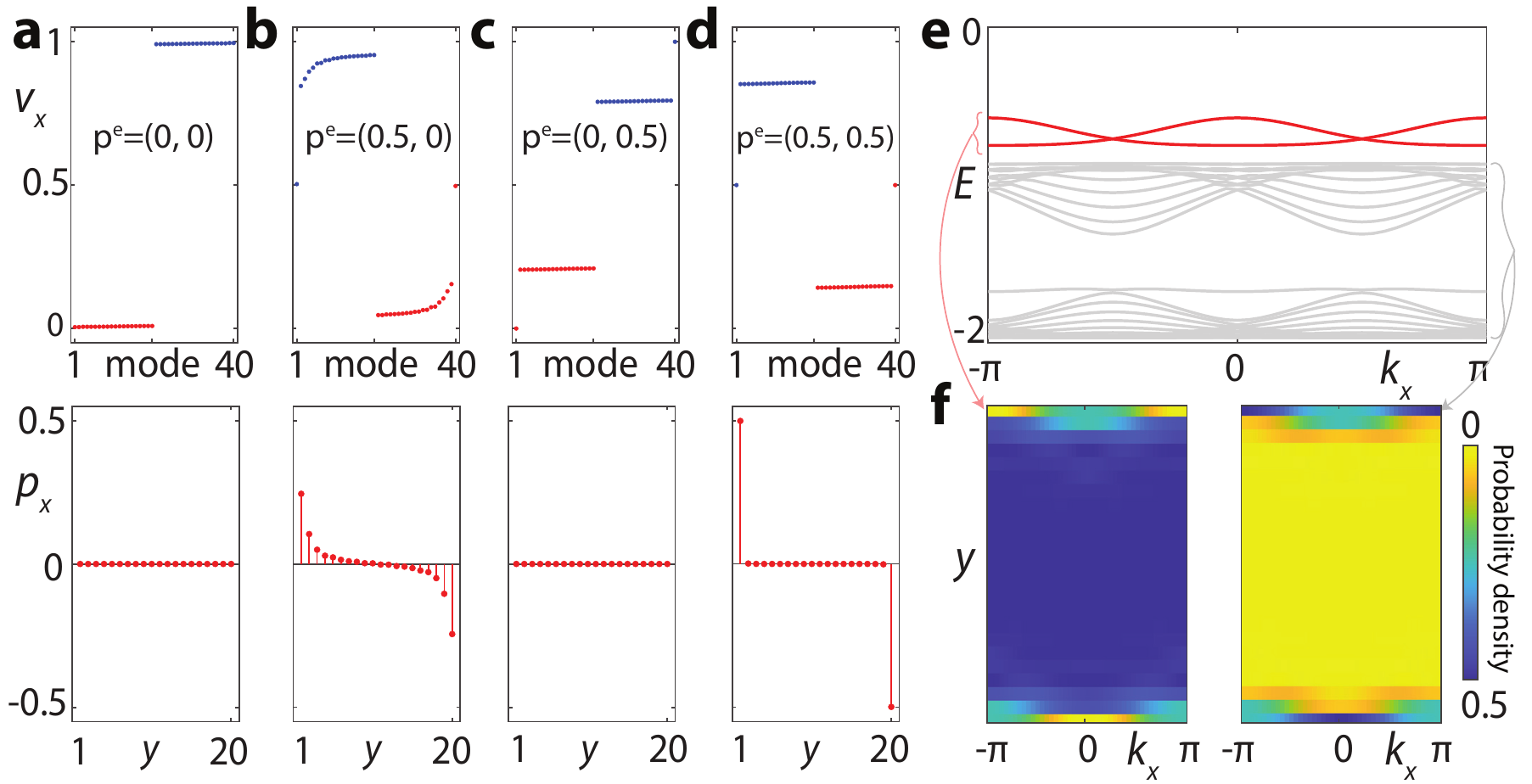}
    \caption{\textbf{ Edge properties.}    
        \textbf{a-d.} Edge Wannier spectra $\{\nu_x\}$ and edge polarization $p_x(R_y)$ are calculated under the parameters $(\gx, \gy)=(1.25, 1.25)$ (a), $(0.25, 1.25)$ (b), $(1.25, 0.25)$ (c), and $(0.25, 0.25)$ (d).          
        \textbf{e-f.} Edge spectrum under $\gx=0.25$, $\gy=0.5$ (e) where densities of edge (left) and bulk (right) bands both exhibit momentum-space glide reflection.
        }
    \label{fig3}
\end{figure}

\textit{Dipole moment quantized by $\kns$ symmetries.}
Although containing momentum half translation, $\kns$ $M_x$, $M_y$, and $P_{\pi}$ quantize electric polarization (see Sec.~V of \cite{SM_note} for detailed derivation):
\begin{subequations}
    \begin{align}
       \{p_x^j(k_y)\} \xlongequal[]{M_x} \{-p_x^j(k_y+\pi)\}~~\mathrm{mod}~\textbf{1}.  \label{eq:mx_polar}  \\
       \{p_y^j(k_x)\} \xlongequal[]{M_y} \{-p_y^j(k_x+\pi)\}~~\mathrm{mod}~\textbf{1}.  \label{eq:my_polar}  \\
       \{p_x^j(k_y)\} \xlongequal[]{P_{\pi}} \{-p_x^j(\pi-k_y)\}~~\mathrm{mod}~\textbf{1}.  \label{eq:p_polarx}  \\
       \{p_y^j(k_x)\} \xlongequal[]{P_{\pi}} \{-p_y^j(\pi-k_x)\}~~\mathrm{mod}~\textbf{1}.  \label{eq:p_polary}
    \end{align}  
    \label{Eq6}
\end{subequations}
The quantization properties above are distinct from those of their non-projective counterparts.
For example, in Eq.~\eqref{eq:mx_polar}, $M_x$ quantizes only the total polarization $p_x$ as 0 or 1/2 but not for each $k_y$, in contrast to the conventional point-group reflection that quantizes polarization for each $k_y$~\cite{alexandradinata2014wilson,benalcazar2017electric}. 
This feature stems from the fact that no wavevector $\textbf{k}$ is invariant under $M_x$, as indicated by the momentum half translation on the right hand of Eq.~\eqref{eq:mx_polar}. 
In addition, $\kns$ $P_{\pi}$ enforces a vanished bulk polarization as the Berry phases of the two filled bands come in pair $(-\nu, \nu)$. 

The system supports quantized edge polarization. Figs.~\ref{fig3}a-d show the Wannier bands (top) and edge polarization $p_x^{\rm e}$ (bottom) of the $y$-ribbon structure (i.e., periodic in $x$ and open in $y$) for $(\gx,\gy)=(0.75\pm0.5,0.75\pm0.5)$, where the edge polarization $p_i^{\rm e}$ is nontrivial for $|\gamma_i|<|\lambda_i|~~(i=x,y)$.
The momentum glide reflection can be validated by the wave function. We compute the edge spectrum of the $y$-ribbon for $\gx=0.25$ and $\gy=0.5$ (Fig.~\ref{fig3}e), where both the charge density of the edge bands (red in Fig.~\ref{fig3}e and Fig.~\ref{fig3}f left) and the bulk bands (gray in Fig.~\ref{fig3}e and Fig.~\ref{fig3}f right) exhibit a half translation along $k_x$, as enforced by the $M_y$ symmetry.

\begin{figure}[htbp]
    \centering
    \includegraphics[width=1\linewidth]{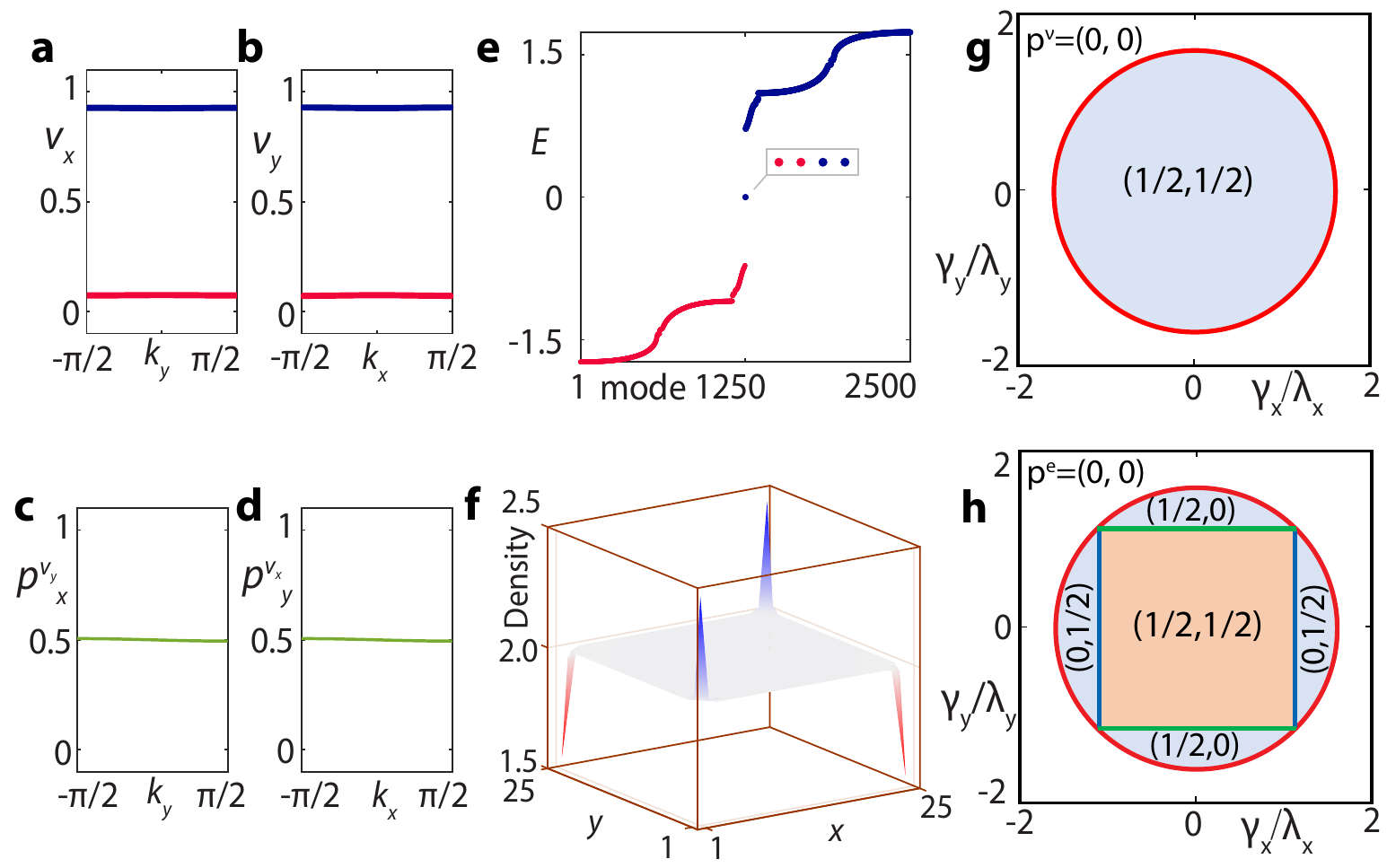}
    \caption{\textbf{Quadrupole insulator on RP$^2$ from $\kns$ symmetries.}
    \textbf{a-b.} Gapped Wannier bands $\nu_x$ and $\nu_y$ showing vanished bulk polarizations.
    \textbf{c-d.} Nontrivial Wannier-sector polarizations.
    \textbf{e.} Appearance of four zero corner modes under full open boundaries for parameters: $\gx=0.25$, $\gy=0.2$, and $N_x=N_y=25$ (number of the unit cell along $x$ and $y$). 
    \textbf{f.} Charge density of corner states (red in Fig.~\ref{fig4}e). 
    \textbf{g.} Isotropic binary bulk phase diagram showing the quadrupole $\mathbf{p}^{\nu}=(1/2,1/2)$ and the trivial phase. Their boundary (red circle) is associated with a bulk gap closing.
    \textbf{h.} Edge polarization $\mathbf{p}^\mathrm{e}$. Only the central square (orange) features simultaneously nontrivial edge polarization $(p^{\mathrm{e}}_x,p^{\mathrm{e}}_y)=(1/2, 1/2)$ and corner modes. The four segments  (light gray regions), although featuring nontrivial bulk quadrupole moments (see g), have nontrivial edge polarization only along a single direction, and corner modes are absent therein.    
    The boundaries between the square and the segments are associated with edge energy gap closing (blue and green lines; see Sec.~VII of \cite{SM_note}).} 
    \label{fig4}
\end{figure}

\textit{Quadrupole moment quantized by $\kns$ symmetries.}
$M_x$ and $M_y$ further quantize the quadrupole moment nonlocally by restricting the Wanner-sector polarization $p_x^{\nu_y}$ and $p_y^{\nu_x}$ as (Sec.~V of \cite{SM_note})
\begin{align}\label{Eq7}
    \{p_x^{\nu_y}(k_y)\} &\xlongequal[]{M_x} \{-p_x^{\nu_y}(k_y+\pi)\}~~\mathrm{mod}~\textbf{1},\\
    \{p_y^{\nu_x}(k_x)\} &\xlongequal[]{M_y} \{-p_y^{\nu_x}(k_x+\pi)\}~~\mathrm{mod}~\textbf{1}.
\end{align}
Whereas $p_x^{\nu_y}$ and $p_y^{\nu_x}$ can be either 0 or 1/2, they must additionally satisfy $p_x^{\nu_y}=p_y^{\nu_x}$ because the phase transition occurs under an isotropic bulk band closing [see Eq.~\eqref{eq:gap} and Fig.~\ref{fig2}e]. This feature differentiates the present model from the originally reported quadrupole model~\cite{benalcazar2017quantized,benalcazar2017electric}, where the phase transition accompanies a Wannier band closing and the Wanner-sector polarizations are independent along different directions.
Shown in Figs.~\ref{fig4}a-b are the gapped Wannier bands symmetric around $\nu_{x,y}=0.5$ in the $\mathrm{RP}^2$ reduced BZ, confirming vanished bulk polarization $p_x=p_y=0$. Such vanished bulk polarization is enforced by $\kns$ symmetry $P_{\pi}$, accordant with the fact that besides being symmetric around $k_i=0~(i=x,y)$, the Wannier band is also symmetric around $k_i=\pm\pi/2~(i=x,y)$ (see Sec. VI of \cite{SM_note}).
Figs.~\ref{fig4}c and d show the Wannier-sector polarizations $p_x^{\nu_y}=p_y^{\nu_x}=0.5$ for the filled Wannier bands (red in Fig.~\ref{fig4}a-b), confirming a nontrivial quadrupole moment. 
Accordingly, the energy spectrum under full open boundaries (Fig.~\ref{fig4}e) contains four degenerate in-gap corner modes at zero energy, and the associated corner charges are quantized as $\pm1/2$ (Fig.~\ref{fig4}f).

Fig.~\ref{fig4}g shows the binary bulk phase diagram, where the nontrivial quadrupole phase region resides inside the red circle and trivial elsewhere. The bulk quadrupole phase transition is triggered by a bulk energy gap closing rather than by Wannier gap closing, indicating a feature of the intrinsic high-order phase~\cite{geier2018second,queiroz2019partial}.
In contrast, in the edge polarization phase diagram (Fig.~\ref{fig4}h), only the central square (orange) shows nontrivial edge polarization along both directions (consequently, zero-energy corner modes under full open boundaries), while the four segments (gray) exhibit nontrivial edge polarization along one and only one direction. The square and the segments are separated by a gap closing (vertical blue and horizontal green lines) in the edge spectrum (see Sec. VII of \cite{SM_note}), a feature of the extrinsic high-order phase where corner modes can be removed without closing the bulk gap~\cite{geier2018second}.
Therefore, the bulk and edge phase diagrams exhibit features of the intrinsic and extrinsic high-order phases, respectively; such a violation of bulk-boundary correspondence relates to the nonsymmorphic nature of the $\kns$ symmetries, which can no longer be obeyed without periodicity.

\begin{figure}
    \centering
    \includegraphics[width=\linewidth]{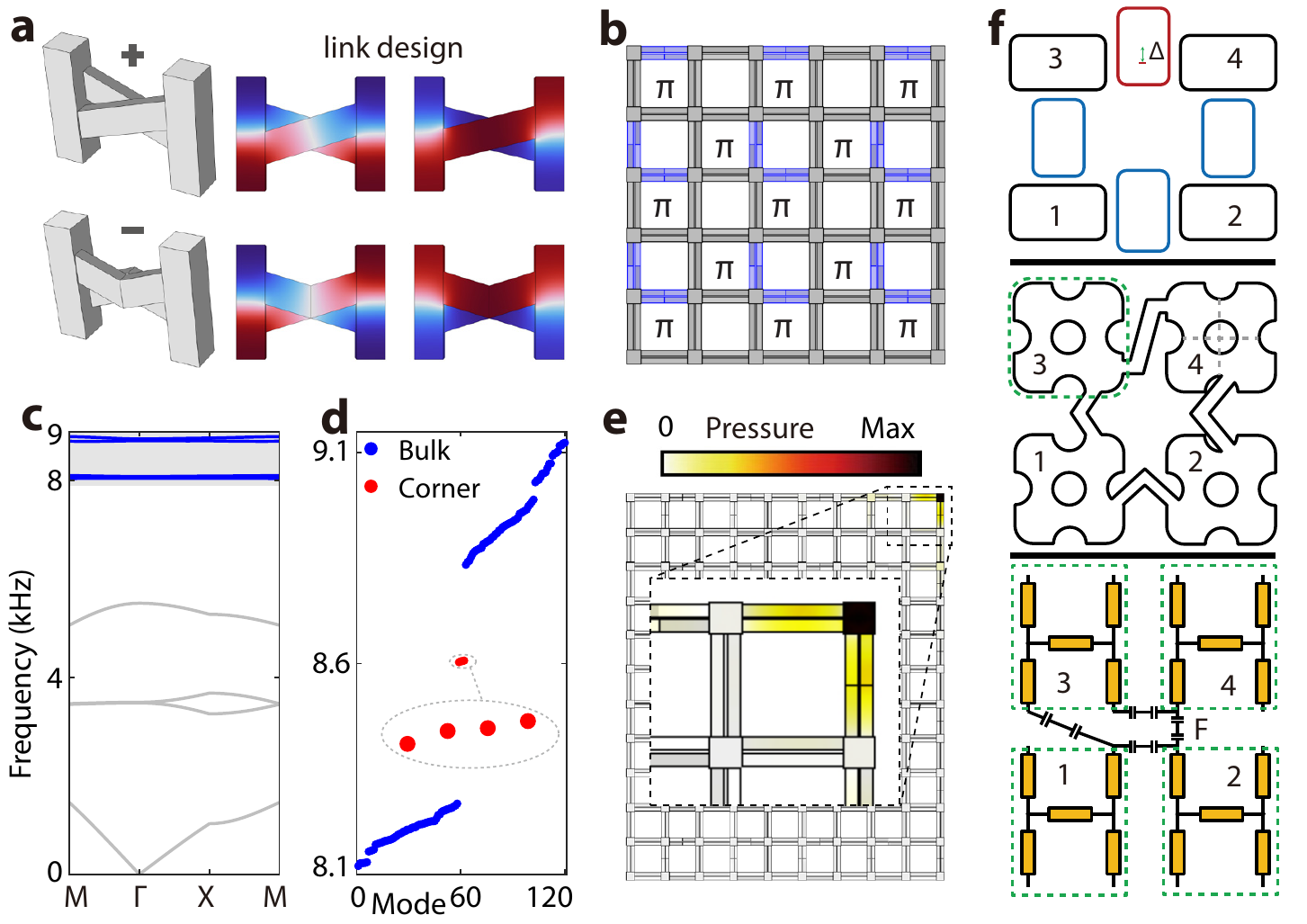}
    \caption{\textbf{Experimental designs.}         
        \textbf{a.} Acoustic links for positive (+) and negative (-) hoppings via V-shaped and cross-shaped connections between two identical resonators, respectively. 
        \textbf{b.} Acoustic resonator array using the link building blocks, where blue color denotes negative coupling. 
        \textbf{c.} Bulk spectrum. The bands of interest are colored in blue, between which a large gap $\approx\SI{700}{\hertz}$ (light gray) appears. 
        \textbf{d.} Eigenfrequencies of an open-boundary structure, exhibiting four in-gap modes. 
        \textbf{e.} Pressure distribution of one in-gap corner mode.
        \textbf{f.} Other feasible platforms: optical ring resonators (top), mechanical metamaterials (middle), and circuits (bottom).        
        }
    \label{fig5}
\end{figure}

\textit{Implementation of Brillouin real projective plane.}
We propose an experimental design to realize the $\mathrm{RP}^2$ BZ and the associated high-order topology based on acoustic resonator arrays~\cite{qi2020acoustic,xue2022projectively,li2022acoustic,xue2020observation,ni2020demonstration}.
The cavity resonator emulating the site of Fig.~\ref{fig2}c is designed to work at the $p_z$ mode where both positive and negative couplings between nearest neighboring resonators are possible~\cite{qi2020acoustic}; these are achieved by linking two identical resonators in the cross and V-shaped manners, respectively (Fig.~\ref{fig5}a). 
In particular, a V-shaped (cross) connected bi-resonator provides the negative (positive) coupling because the in-phase coupled mode is at a lower (higher) frequency, while the out-of-phase coupled mode is at a higher (lower) frequency (see Sec.~VIII of \cite{SM_note}). 
The coupling strength is strongly related to the tube cross-section and tube position with respect to the middle of the resonator. 
Without loss of generality, we fix the tube thickness and scan the tube width and position, from which distinct coupling strengths can be designed. To get a wide band gap, a large coupling contrast is utilized between intra-cell and inter-cell hoppings that are \SI{47}{\hertz} and \SI{482.2}{\hertz}, respectively. 

Based on the link building blocks, we design the model (Fig.~\ref{fig2}c) by constructing an acoustic resonator array featuring the checkerboard $\pi$ flux configuration in Fig.~\ref{fig5}b, where the blue color indicates negative coupling. The bulk energy spectrum is present in Fig.~\ref{fig5}c, where the bands of interest (blue) are separated by a large gap $\approx\SI{700}{\hertz}$ (light gray). We computed the eigenfrequency spectrum for an array of $5\times6$ cells under the open boundary condition. The results are shown in Fig.~\ref{fig5}d, which reveals the presence of four in-gap corner modes, as verified by their corresponding sound pressure distributions located at the corners of the array (Fig.~\ref{fig5}e).
The same $\mathrm{RP}^2$-BZ model can be realized with coupled optical ring resonators (Fig.~\ref{fig5}f top), mechanical metamaterials (Fig.~\ref{fig5}f middle), and circuits (Fig.~\ref{fig5}f bottom); we provide more detailed discussions on these platforms in Sec.~IX of~\cite{SM_note}. 

\indent In summary, we report the construction of a Brillouin real projective plane by means of momentum-space nonsymmorphic ($\kns$) symmetries that are enabled by $\mathbb{Z}_2$ synthetic gauge fields. Theoretical derivation reveals that $\kns$ symmetries quantize bulk polarization and Wannier-sector polarization nonlocally across different momenta, giving rise to a high-order quadrupole phase with a phase transition of bulk energy gap closing. 
At last, we show that the Brillouin real projective plane and its quadrupole phase can be viably realized on many engineered physical platforms.

\textit{Note added.} During the completion of this work about high-order topology on Brillouin $\mathrm{RP}^2$, we noticed three relevant recent preprints~\cite{wang2023chessboard,zhu2023brillouin,tao2023higherorder}, which used momentum-space nonsymmorphic symmetries for first-order and high-order topology in two- and three-dimensional Brillouin Klein bottles. 
\bibliographystyle{apsrev4-2}
\bibliography{manuscript}
\end{document}